\newcommand{\ms}[1]{\mathsf{#1}} 

\documentclass[conference]{IEEEtran}

\usepackage{amsmath}
\usepackage{amsfonts}
\usepackage{multirow}
\usepackage{graphicx}
\usepackage{algorithm}
\usepackage{algorithmic}

\usepackage{svg}
\usepackage{tabularx}
\usepackage[utf8]{inputenc}
\usepackage{soul}
\usepackage{xcolor}
\usepackage{nicefrac}
\usepackage{float}

\usepackage[pdfstartview=XYZ,
bookmarks=true,
colorlinks=true,
linkcolor=blue,
urlcolor=blue,
citecolor=blue,
pdftex,
bookmarks=true,
linktocpage=true, 
hyperindex=true
]{hyperref}
\usepackage{cleveref}

\crefname{table}{Table-}{Tables-}
\Crefname{table}{Table-}{Tables-}
\usepackage{balance}

\graphicspath{{figures/}}%

\begin{document}
\title{if-ZKP: Intel FPGA-Based Acceleration of Zero Knowledge Proofs }
 \author{
     \IEEEauthorblockN{
     Shahzad Ahmad Butt\IEEEauthorrefmark{1}\textsuperscript{\dag},
     Benjamin Reynolds\IEEEauthorrefmark{1},
     Veeraraghavan Ramamurthy\IEEEauthorrefmark{1},\\
     Xiao Xiao\IEEEauthorrefmark{1},
     Pohrong Chu\IEEEauthorrefmark{1},
     Setareh Sharifian\IEEEauthorrefmark{1},
     Sergey Gribok\IEEEauthorrefmark{1},
     Bogdan Pasca\IEEEauthorrefmark{1}
     }

     \IEEEauthorblockA{\IEEEauthorrefmark{1}Intel Corporation\\   
     101 Innovation Dr, San Jose, USA\\
     Email: \{first.lastname\}@intel.com\textsuperscript{*}\\
     \textsuperscript{\dag}Shahzad Ahmad Butt's email is shahzad.ahmad.butt@intel.com.}
 }
\maketitle
                                                                              
\begin{abstract}
Zero-Knowledge Proofs (ZKPs) have emerged as an important cryptographic technique allowing one party (prover) to prove the correctness of a statement to some other party (verifier)
and nothing else. ZKPs give rise to user's privacy in many
applications such as blockchains,
digital voting, and machine learning. Traditionally, ZKPs suffered from poor scalability but recently, a sub-class of ZKPs known as Zero-knowledge Succinct Non-interactive ARgument of Knowledges (zk-SNARKs) have addressed this challenge.  They are getting significant attention and are being implemented by many public libraries.
In this paper, we present a novel scalable architecture
that is suitable for accelerating the zk-SNARK prover compute on FPGAs. We focus on the multi-scalar multiplication (MSM) that accounts for the majority of computation time spent in zk-SNARK systems. The MSM calculations extensive rely on modular arithmetic so highly optimized Intel IP Libraries for modular arithmetic are used. The proposed architecture exploits the parallelism inherent to MSM and is implemented using the Intel OneAPI framework for FPGAs. Our implementation runs 110x-150x faster compared to reference software library, uses a generic curve form in Jacobian coordinates and is the first to report FPGA hardware acceleration results for BLS12-381 and BN128 family of elliptic curves.  
\end{abstract}
\section{Introduction}
As our dependence on online activities continues to increase, the presence of the Internet brings forth potential risks. Our vulnerability to scams, identity theft, intrusive advertisers, governmental control, and unidentified sources of danger also grows simultaneously. Privacy-enhancing technologies (PETs) serve as a remedy for these privacy risks by granting users the ability to regulate the gathering, dissemination, and utilization of personal information and daily actions. These sophisticated PETs employ intricate cryptographic methods to offer privacy-focused alternatives to technologies that neglect the importance of safeguarding personal privacy.

There are many PETs which may considered as part of data protection
compliance including Homomorphic encryption (HE), Secure multiparty computation (SMPC), Trusted execution environments (TEE) and Zero-knowledge proofs (ZKP).
In this paper we focus on ZKP due to it's growing prominence in blockchain projects such as Ethereum, Filecoin and Zcash. Fundamentally, ZKPs enable a \emph{prover} to assure a \emph{verifier(s)} of the truthfulness of a statement without divulging any confidential information and do so in a secure fashion. This dynamic allows a  \emph{verifier} to authenticate information while simultaneously enabling the \emph{prover} to maintain the confidentiality of their secret data. The potential of this relationship for enhancing security systems across various applications is significant provided they can be feasibly implemented in real-world scenarios.  
This recent increase in attention is due to the advancement of a ZKP type known as a zk-SNARK. More precisely, zk-SNARK constructions such as Groth16 \cite{groth2016size} and PlonK \cite{gabizon2019plonk} have enabled significant progress in our ability to develop functional zk-SNARKs capable of deployment in production level systems. 
 
A closer examination of these zk-SNARK systems reveals that although the proofs outputs are small for quick verification, the proof generation process itself is notably computationally intensive. This is because to uphold stringent security standards in these schemes, these systems consist of tens or hundreds of millions of constraints and utilize technology such as Elliptic Curve Cryptography (ECC) which requires many large bitwidth mathematical operations. For example, in applications such as Filecoin, the proof circuits used in their SNARKs encompass up to $2^{27}$ constraints~\cite{protocolaiSnarks_filecoin_circuit_size} and operate on 381-bitwidth numbers. This level of complexity underscores the immense compute requirement for secure and reliable proof generation in these systems. 
 
For this reason, both industry and academia have started to develop solutions to offload this compute to accelerators, aiming to enhance the practicality of ZKP schemes being deployed in real world applications. While GPUs, due to their robust software ecosystem and established presence in the blockchain industry, are often the primary target for acceleration solutions in this domain, we believe that FPGAs provide the flexibility and capability for hardware fine-tuning which is essential to building an efficient solution in the rapidly evolving blockchain industry.

Our benchmarking experiments, along with those conducted by others, on the proof generation process in functional implementations of zk-SNARK schemes such as Groth16, have highlighted the primary targets for acceleration offload to be the mathematical operations Multi-scalar Multiplication (MSM) and Number Theoretic Transform (NTT). 
Competitions such as ZPrize~\cite{Zprize_2022_github} have played a pivotal role in advancing both GPU and FPGA participation of acceleration of MSM and NTT functions in an open-source setting, but have focused on elliptic curves which benefit from what is known as Twisted Edwards representation, such as BLS12-377. This representation notably reduces the mathematical compute requirements on hardware which mitigates much of the area consumption challenges that come with building a solution in this domain at the cost of overall system security. In contrast, our work aims to tackle what we believe to be more general and relevant curves in BN128 and BLS12-381 which must be represented in Weierstrass form and pose larger design challenges but also come with greater industry adoption.  

In this paper, our primary focus is on the Multi-scalar Multiplication (MSM) mathematical function, a critical component in zk-SNARK solutions that often creates performance bottlenecks. Our objective is to address these computational challenges through an optimized solution developed for FPGAs, with a particular emphasis on the Intel Agilex architecture. 
The key contributions of our work are outlined as follows:
\begin{itemize}
    \item \textbf{Novel Scalable Architecture for Bucket Algorithm:} This work presents a scalable architecture to accelerate MSM calculations
    \item \textbf{Use of Recursive Bucket Method:} This work harnesses the bucket method in a recursive fashion to minimize the computational requirements during the combination phase of the algorithm.
    \item \textbf{Unified Double and Add Unit:} The paper explores the design of a fully pipelined point double and point add unit that can run at very high frequency compared to other reported results in the literature   
    \item \textbf{Use of non-Montgomery/Standard form numbers:} This work has explored the implementation of Elliptic Curve Point Addition and Doubling using standard/integer format numbers to improve quality of results and fit within FPGA resource limitations.
    \item \textbf{First Implementation of BLS12-381 on FPGA:} Fully functional implementation of MSM for BLS12-381 curve validated on Intel Agilex FPGA.
\end{itemize}

\section{Preliminaries}

\subsection{Proof Systems in Cryptography}
 A \textbf{proof system} in the cryptographical sense, is any procedure by which one party, referred to as the \textit{prover} $\ms{P}$,
wishes to convince another party,  referred to as the \textit{verifier} $\ms{V}$, that a given statement is true. 
A proof system should be \textit{complete} and \textit{sound}. \textit{Completeness} means any true statement should have a convincing proof of its validity and \textit{soundness} means no false statement should have a convincing proof. 

A \textbf{knowledge proof} goes beyond simply proving the correctness of a statement, as it also demonstrates that the prover possesses \textit{knowledge} of  some witness $w$ that satisfies the statement. We define an efficient algorithm $\ms{A}$ to \textit{know} a value $w$ if it is possible to construct another efficient algorithm that takes $\ms{A}$ as input and outputs $w$. This algorithm is referred to as an \textit{extractor} for $\ms{A}$.

In a \textbf{zero-knowledge proof (ZKP)}, we also require the proof to reveal nothing more than the truth of the statement. 
In a ZKP, the (potentially dishonest) verifier $\ms V$, does not acquire any new knowledge from their interaction with the prover $\ms P$ on a common input $x$, and all the information that $\ms V$ gets can be derived directly from $x$ through an efficient algorithm.

\subsection{zk-SNARKs}
A growing real-world application of ZKPs is in blockchain projects
such as Ethereum, Filecoin and Zcash. The particular type of zero-knowledge arguments deployed in these blockchains is a (pre-processing) succinct non-interactive argument of knowledge, or
zk-SNARK for short. 
The term \textit{succinct} indicates that the proof's size is significantly smaller than the statement or witness size, facilitating quick verification. The \textit{Non-Interactive} part means the communication is one-way and there is no need for multiple interactions between the prover and the verifier. The last part is the \textit{Argument of Knowledge} which means a computationally bounded prover cannot construct a proof without knowing a certain so-called witness for the statement. However, it is knowledge-sound {only} against provers with bounded computational resources.

Compared to other forms of zero-knowledge arguments, zk-SNARKs are particularly well-suited for the blockchain environment. This is due to the fact that proofs can be easily stored, accessed, and verified on a publicly accessible blockchain.

The security of some of zk-SNARKS relies on variants of the \textit{discrete logarithm problem (DLP)}.
Hyrax \cite{wahby2018doubly} is an example of such schemes. The \textit{bilinear pairing} is one of the techniques used for constructing zk-SNARKs. It is used in schemes such as Groth16 \cite{groth2016size}, Plonk \cite{gabizon2019plonk} and Sonic \cite{maller2019sonic}. In modern cryptography, the groups used for DLP-based zk-SNARKs are typically cyclic subgroups of groups defined via {elliptic curves}
over finite fields, or the multiplicative group of integers modulo a very large prime $p$. For pairing-based zk-SNARKs, the only known instantiation  is over dedicated algebraic elliptic curves named
pairing-friendly, and the most efficient ones are specific elliptic curves named 
after Barreto, Lynn and Scott (BLS \cite{barreto2003constructing}), and Barreto–Naehrig (BN \cite{barreto2005pairing}).

{\color{black}
\subsection{Elliptic Curves Groups and their Basic Operations}
Elliptic curves have long been recognised by number theorists as a generalization of the multiplicative group. This field is extensively explored in discrete mathematics. Here, we briefly review the basic elliptic curves notions and operations.    

An \textit{elliptic curve group} is established in relation to a finite field $\mathbb{F}$, referred to as the curve's base field. Group elements represent pairs of coordinates $(x, y)\in \mathbb{F}\times\mathbb{F}$ satisfying an equation of the form $y^2 = x^3 + ax + b$, where $a$ and $b$ are designated field elements. This is the general form of elliptic curves and is referred to as the \textit{``Weierstrass form''}.

Points \( P \) and \( Q \) on this curve can be added together to produce another point \( R \). This operation is associative and has an identity element (the point at infinity, denoted by \( O \)), making the set of points on the curve, together with the addition operation, form an Abelian group.

The \textit{addition} of two points $P=(x_1,y_1)$ and $Q=(x_2,y_2)$ is defined using one of the following three rules:
\begin{itemize}
\item if $x_1 = x_2$ and $y_1 =-y_2$, define $P+Q := O$.
\item if $P=Q$, 
find the \textit{tangent line} crossing point $P$ and then find the line's intersection with the curve at point $R=(x,y)$. Then, $P+P=2P:=(x,-y)$. The slope of the tangent line is $s_t=\frac{3x_1^2+a}{2y_1}$.
\item if none of the above, 
draw the \textit{chord line}  through the two points which intersects the elliptic curve at a third
point $R = (x, y)$. Then,  $P+Q:=(x,-y)$. The slope of the line between these two points is $s_c=\frac{y_1-y_2}{x_1-x_2}$
\end{itemize}

\textit{Scalar multiplication} for curve points is another important operation. Given a scalar \( s \) and a point \( P \), the scalar multiplication \( s \cdot P \) is defined as adding point $P$ for $s$ times.
For large values of \( s \), directly adding \( P \) to itself \( s \) times is inefficient. Therefore, expediting methods like double-and-add are used.

\noindent \textbf{Edwards Curves.} An Edwards curve over a non-binary field $\mathbb{F}$ is a curve $E: x^2+y^2 =1+dx^2y^2$ where $d\in \mathbb{F}\setminus\{0, 1\}$. 
This curve can be put into Weierstrass form via a simple
rational change of variable. 
The addition law is strongly unified: i.e., it can also be used to double a point.

\noindent \textbf{Twisted Edward Curves.} 
Twisted Edwards curves are embedded  set of Edwards
curves in a larger set of elliptic curves of a similar shape. More specifically, the twisted Edwards curve over $\mathbb{F}$ with coefficients $a, d \in \mathbb{F}$  is the curve $TE: ax^2 + y^2 = 1 + dx^2y^2$. As observed, an Edwards curve is a twisted Edwards curve with $a = 1$.
}

Group operations are more efficient than other forms of elliptic curves for Edwards curves \cite{bernstein2007faster} and, more generally, twisted Edwards curves \cite{bernstein2008twisted}. This is because addition formulas are \textit{complete}. in such curves. Meaning that the formulas work for all
pairs of input points on the curve, with no exceptions for doubling or neutral element. 
\subsection{The Computational Challenge: Complex Operations}

Setting-up and using a zk-SNARK involves complex mathematics, but runtime analysis shows that only a few compute-intensive components are used extensively. As it currently stands, for the state-of-the-art in zk-SNARKs these components are well known \cite{zhang2021pipezk_fpga, aasaraai2022cyclonemsm_fpga, lu2022cuzk_gpu, ma2023gzkp_gpu, ni2023enabling_ntt_gpu, vezenov2022accelerating_gpu_ntt_msm, derei2023accelerating_plonk_msm_ntt_gpu} 
and experts have a consensus on it. For our analysis and verification of compute-intensive components, we did software profiling on the open-source implementation \emph{libsnark} \cite{libsnark} to profile the \emph{prover algorithm}.  The results shown in Table~\ref{tab:profile_prov_bn128_bls12_381} clearly show that most of the computation revolves around three core operations namely, \emph{Multi-Scalar Multiplication (MSM) on $\mathbb{G}_1$} elliptic curve points, \emph{MSM on $\mathbb{G}_2$} elliptic curve points and \emph{Number-Theoretic Transform (NTT)} (both forward and reverse directions). The MSM on $\mathbb{G}_1$  and $\mathbb{G}_2$  elliptic curve points share the exact same compute algorithm and arithmetic operations but differ in the representation of points in $\mathbb{G}_1$ and $\mathbb{G}_2$  groups. From this analysis, the first candidate we chose for accelerating the zk-SNARKs prover algorithm is the \emph{MSM on $\mathbb{G}_1$ } points --  targeted in this work -- leaving the NTT and the adaptation for $\mathbb{G}_2$  MSM as future works.

\begin{table}
\begin{footnotesize}
\begin{center}
\caption{\label{tab:profile_prov_bn128_bls12_381} Prover Profiling Results for the BN128 and BLS12-381 Curves}
\begin{tabular}{c || c c c c}
\hline
\multirow{2}{*}{\bf Curve}    & \multicolumn{4}{c}{\bf Operation} \\
           \cline{2-5}
           &  \bf MSM-$\mathbb{G}_1$ & \bf MSM-$\mathbb{G}_2$ & \bf NTT & \bf Other \\  
\hline \hline
\bf BN128      & 37\%  & \bf 51\%   & 11\% & 1\%  \\
\bf BLS12-381  & 33\%  & \bf 59\%   & 7\%  & 1\%  \\
\hline
\end{tabular}
\end{center}
\end{footnotesize}
\end{table}

\subsection{Multi-Scalar Multiplication (MSM)}
Multi-scalar multiplication (MSM) is an essential operation in many public-key cryptosystems. It involves the simultaneous multiplication of multiple scalars with different elliptic curve points. It is a dot product operation between two vectors of different types, one being a set of scalars within a prime field and the other being a set of points on an elliptic curve.
Given scalars \(\mathcal{S}=\{ s_1, s_2, \ldots, s_m \}\) and points \(\mathcal{P}=\{P_1, P_2, \ldots, P_m \}\) on an elliptic curve, the multi-scalar multiplication is defined as:
\vspace{-0.8em}
\[
R = \sum_{i=1}^{m} s_i  P_i,
\]
where each \( s_i P_i \) denotes the scalar multiplication of point \( P_i \) with scalar \( s_i \).
The most ordinary and obvious way to evaluate this dot product is by multiplying corresponding scalars and points followed by accumulation. 
The challenge with multi-scalar multiplication is to perform the operation more efficiently than simply computing each scalar multiplication separately and then summing the results.

For instance, given scalars \( s_1 \) and \( s_2 \) and points \( P_1 \) and \( P_2 \), the naive method would compute $
R_1 = s_1 \cdot P_1$, $R_2 = s_2 \cdot P_2$ and then 
$R = R_1 + R_2$. 
However, efficient algorithms like Shamir's Trick or Pippenger's method (Bucket Algorithm) can perform multi-scalar multiplication faster than the naive method, especially when \( m \), 
is large.

The known way to multiply scalars and elliptic curve points relies on what is known as the \emph{double-and-add algorithm} and is presented in Algorithm~\ref{algo:double_and_add} for a scalar $s$ and a point $P$.

\begin{algorithm}[tb]
\caption{\emph{Double-and-Add Algorithm} for Elliptic Curve Scalar Multiplication}
\label{algo:double_and_add}
\begin{footnotesize}
\begin{algorithmic}
        \STATE \textbf{Input:} $s$ \hspace{0.5in} \textit{\textcolor{gray}{// Scalar}} 
        \STATE \textbf{Input:} $P$ \hspace{0.49in}\textit{\textcolor{gray}{// Point on the elliptic curve}}
        \STATE \textbf{Output:} $Q=sP$ \hspace{0.10in}\textit{\textcolor{gray}{// The product}}

\STATE $Q \leftarrow O$ \hspace{0.55in}\textit{\textcolor{gray}{// Initialize \( Q \) as the point at infinity (\( O \))}}
\STATE \hspace{0.01in}\textit{\textcolor{gray}{// Iterate over the bits of $s$}}
\FOR{\( s[j] \) in \( s \), from \textit{MSB} down to \textit{LSB}} 
\STATE \( Q = 2Q \) \hspace{0.34in} \textit{\textcolor{gray}{// Doubling step}} 
\IF{\( s[j] = 1 \)}
\STATE \( Q = Q + P \) \hspace{0.02in} \textit{\textcolor{gray}{// Addition step}}  
\ENDIF
\ENDFOR
\RETURN \( Q \)
\end{algorithmic}
\end{footnotesize}
\end{algorithm}

Algorithm~\ref{algo:double_and_add} has a complexity of of $\mathcal{O}(N)$ where $N$ is the bitwidth of scalars. What this complexity entails is the number of elliptic curve additions or doubling operations. To clarify the context, one needs to understand what is the nominal value of $N$ and how complex the addition or doubling operations are for elliptic curve points. In this work, we have used BN128 and BLS12-381 curves \cite{bn128_curve_standard_draft, BarretoNaehrig2006_bls12_381_curve_ref_bn128_ref} for which the scalar widths $N$ are 254 and 381 bits respectively. The \emph{double-and-add} operation on elliptic curve points is a very complex mathematical operation and can involve between 8 to 16 modular multiplications where the size of multiplication inputs in bits is $254\times254$ for BN128 and $381\times381$ BLS12-381. The amount of compute to be performed for calculating the size-$m$ MSM using this algorithm is shown in Table~\ref{table:msm_multiplication}.

\begin{table}
    \centering
    \caption{Number of Modular Multiplications for calculating MSM using Double-and-Add Algorithm}
    \label{table:msm_multiplication}
    \begin{tabular}{|c|c|}
        \hline
        \textbf{Curve} & \textbf{Number of Modular Multiplications} \\
        \hline
        BN128 & \( m \times (2 \times 254 \times 16) \) \\
        \hline
        BLS12\_381 & \( m \times (2 \times 381 \times 16) \) \\
        \hline
    \end{tabular}
\end{table}

Since the size of the MSM ($m$) for the problems of interest nowadays \cite{protocolaiSnarks_filecoin_circuit_size} can exceed 100 million points -- a better algorithm is required for MSM. The \emph{Bucket algorithm} \cite{pippenger1976evaluation} is a well known optimized way of calculating MSM particularly interesting for large values of $m$.

\subsection{Bucket Algorithm} 
The \emph{Bucket Algorithm} describes an optimized approach to calculate MSM when the scalar bitwidth $N$ is small. 
Most of the MSM calculations for zk-SNARK applications involve scalars with very wide bitwidths, typically ranging from 254 bits and extending to 700 bits or beyond. The standard way of using the bucket algorithm is to first refactor the original MSM computation into many new MSM computations with the same size but operating on narrow-bitwidth scalars and finally combine them to calculate the required result.

Let $s_i$ be an $N$-bit scalar and assume for simplicity that $N$ is a multiple of $p$. We define: 
$k=\frac{N}{p}$
as the limbs bitwidth of $s_{i,j}$ with $j\in[0,p-1]$:
$$
s_i = s_{i,p-1}2^{k(p-1)}+\dots+s_{i,1}2^k+s_{i,0}
$$

Let $\mathcal{S} = \{s_1, s_2,\dots,s_m\}$ be a length $m$ array of containing scalars.
The array may be partitioned into $p$ sub-arrays, each of $m$ elements $\{ \mathcal{S}_0, \mathcal{S}_2, \dots, \mathcal{S}_{p-1} \}$,
where each $\mathcal{S}_j$, $j\in[0, p-1]$ contains slice index $j$ of the corresponding $s_{i,j}$ scalars:
$$
\mathcal{S}_i = \{ s_{1,i}, s_{2,i}, \dots, s_{m,i}\}.
$$

Next, we compute $p$ new MSMs and combine them in order to obtain the final MSM output:
\[ \text{MSM}(\mathcal{S},\mathcal{P}) = \text{Comb}\{ \text{MSM}_0(\mathcal{S}_0,\mathcal{P}), \dots, \text{MSM}_{p-1}(\mathcal{S}_{p-1},\mathcal{P}) \}. \]
Each \emph{small} $\textrm{MSM}_j$, $j\in[0,p-1]$ is computed as:
$$
\text{MSM}_j(\mathcal{S}_j,\mathcal{P}) = \sum_{i=1}^{m} s_{i,j} P_i. 
$$
These are combined to compute the final result using the \emph{double-and-add} routine given below:
\[
\text{MSM} = \sum_{j=0}^{p-1} 2^{kj} \text{MSM}_j
\]

The smaller MSMs are calculated using the bucket method which is given in Algorithm~\ref{alg:bucket_msm}.

Table~\ref{table:msm_bucket_compute_complexity} illustrates the compute reduction achieved when using the bucket algorithm as opposed to the double-and-add method described earlier. These results were obtained by employing 
tailored parameters optimized for hardware implementation on the Intel Agilex FPGA 
(our target device).

\begin{table}
    \centering
    \caption{Number of Modular Multiplications and Compute Reduction using Pippenger's Algorithm}
    \label{table:msm_bucket_compute_complexity}
    \begin{tabular}{|c|c|c|}
        \hline
        \textbf{Curve} & \textbf{Number of Modular Multiplications} & \textbf{Reduction} \\
        \hline
        BN128 & \( m \times 22 \) & \( 23\times \) \\
        \hline
        BL12-381 & \( m \times 32 \) & \( 24\times \) \\
        \hline
    \end{tabular}
\end{table}

\begin{algorithm}[tb]
\caption{Bucket Algorithm for MSM}
\label{alg:bucket_msm}
\begin{footnotesize}
\begin{algorithmic}
\STATE \textbf{Input:} $k$ \hspace{1.07in} \textit{\textcolor{gray}{// Scalar bitwidth}} 
\STATE \textbf{Input:} \{$s_1,\hdots,s_m$\} \hspace{0.5in} \textit{\textcolor{gray}{// Array of $m$ scalars}} 
\STATE \textbf{Input:} \{$P_1,\hdots,P_m$\} \hspace{0.49in}\textit{\textcolor{gray}{// Array of elliptic curve points}}
\STATE \textbf{Output:} $R=\sum_{i=1}^{m}s_iP_i$ \hspace{0.32in}\textit{\textcolor{gray}{// The multi-scalar multiplication result}}

\textit{\textcolor{gray}{// Initialize array with zero (points at infinity)}}
\STATE $B=\{O,\hdots,O$\} \hspace{0.72in}\textit{\textcolor{gray}{// Array of size $2^k$}} 
\FOR{$i = 1$ to $m$}
\STATE $B[s_i] = B[s_i] + P_i$ \hspace{0.46in}\textit{\textcolor{gray}{// Fill buckets}}
\ENDFOR

\hspace{0.0in}\textit{\textcolor{gray}{// Reconstruct MSM result form bucket data}}
\STATE $A = 0$          \hspace{0.46in}\textit{\textcolor{gray}{// Initialize accumulator at 0 (elliptic curve type)}}
\STATE $E = B[2^{k}-1]$  \hspace{0.00in} \textit{\textcolor{gray}{// Initialize base with top bucket value}}
\FOR{$i = 2^{k-1}$ to $1$}
\STATE $A = A + E$
\STATE $E = E + B[i-1]$
\ENDFOR
\RETURN $A$ \hspace{0.30in}\textit{\textcolor{gray}{// Return the multi-scalar multiplication result}}
\end{algorithmic}
\end{footnotesize}
\end{algorithm}

\section{Related Works}

In the realm of zk-SNARKs, it's noteworthy that only within the last decade have these schemes been refined and made accessible to the public in a usable form such as open-source libraries like libsnark\cite{libsnark} or Bellman\cite{bellman}. Recently, 
research and competitions have emerged \cite{Zprize_2022_github}   geared towards developing accelerators for enhancing the efficiency and real-world performance of zk-SNARK systems through ASIC, GPU, or FPGA acceleration of the most compute-intensive components.  
 
PipeZK\cite{zhang2021pipezk_fpga} was a recent work that helped introduce zk-SNARK hardware acceleration using an ASIC design method.  Although it works well in small-scale applications, it didn’t scale well and the performance degraded with large-scale applications where large on-chip storage is needed.
  
CuZK \cite{lu2022cuzk_gpu} introduced an innovative method for processing MSM by converting the serial Pippenger algorithm to a sequence of sparse matrix operations, achieving parallelism by deploying multiple threads on the GPU. This resulted in a speed-up that scales linearly with number of the threads.  
While their work provided a novel approach for accelerating BLS12-381 on a GPU, it lacked a comparative analysis against an optimized FPGA implementation for that same curve and did not provide power consumption measurements for efficiency analysis. 

GZKP\cite{ma2023gzkp_gpu} is another GPU-based implementation that aims to showcase the capabilities of its optimized finite field library, which supports arithmetic operations on integers as large as 753-bits. Furthermore, GZKP builds a MSM module that exploits redundant elliptic curve operations as a technique to reduce overall compute and in-turn accelerates the overall MSM calculation.
 
In addition to GPU and ASIC, we have seen some designs utilize FPGA to accomplish the acceleration.  Notably, CycloneMSM \cite{aasaraai2022cyclonemsm_fpga} and ZPrize\cite{Zprize_2022_github} have provided innovative implementations of MSM tailored for accelerating the BLS12-377 curve. These solutions proved the viability of using FPGAs to accelerate MSM calculation, however they have utilized the Twisted Edwards elliptic curve form which has drawback that not every curve can be transformed to it. Our solution uses a generalized elliptic curve form in Jacobian coordinates.

Our implementations followed a two prong meet in the middle approach with focus on developing an optimized scalable architecture suitable for FPGA based accelerators and the design of highly optimized elliptic curve point processor meticulously optimized for the Intel Agilex FPGA architecture. Having built such an optimized point processor allowed us to be the first to support BLS12-381 curve on FPGA.

\section{Hardware Architecture} \label{sec:hardware_architecture}
The fundamental mathematical computation that is carried out while performing MSM consists only of two core operations namely: {\bf Elliptic Curve Point Addition} and {\bf Elliptic Curve Point Doubling}.
Most of the compute blocks used within the naive implementation of MSM or even optimized implementation based on Bucket Method~\cite{pippenger1976evaluation} solely rely on these two operations. Depending on the coordinate system used for curve points there are many different formulas for point addition and point doubling. Our implementation uses Jacobian coordinates where the cost of calculating a single point addition in hardware is 16 modulo multiplications and point double requires 9. The size of each multiplication inputs being 254\texttimes254 or 381\texttimes381 bits. Even when different point coordinate systems, combined with some tricky form of the elliptic curve are used to optimize point add/double compute the required number of hardware resources for high throughput pipelined implementation are extremely high. In our implementation more than 30\% of FPGA fabric on a large device is getting used for just a single instance of point adder. Due to this reason, its a consensus in most of the implementations \cite{zhang2021pipezk_fpga,aasaraai2022cyclonemsm_fpga} reported so far that MSM on an FPGA should be built around a single or few point add unit instances that can be shared between all other algorithmic blocks.

\begin{figure}
  \centering
  \includegraphics[width=.65\linewidth]{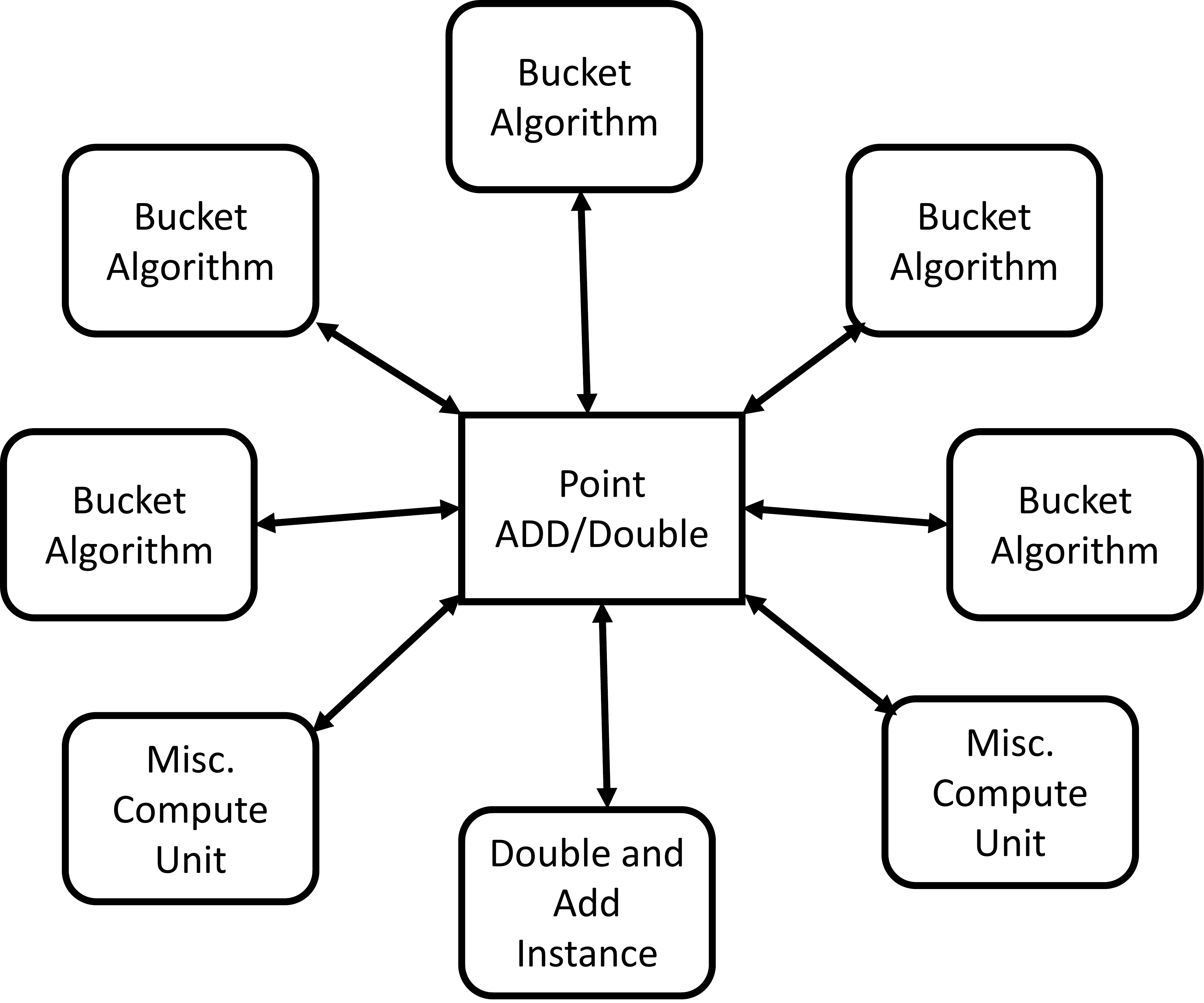} 
  \caption{Sharing of Elliptic Curve Add/Double unit for Compute}
  \label{fig:shared_pa_pd} 
\vskip -0.4cm
\end{figure}

Figure~\ref{fig:shared_pa_pd} shows the description of an important architectural paradigm elaborating how the functional compute units namely point add and point double get shared between different algorithmic compute blocks that carry out different tasks to calculate the final MSM.

\subsection{Scalable Architecture for Bucket Algorithm (SAB)}
\begin{figure}[h] 
  \centering
  \includegraphics[width=.7\linewidth]{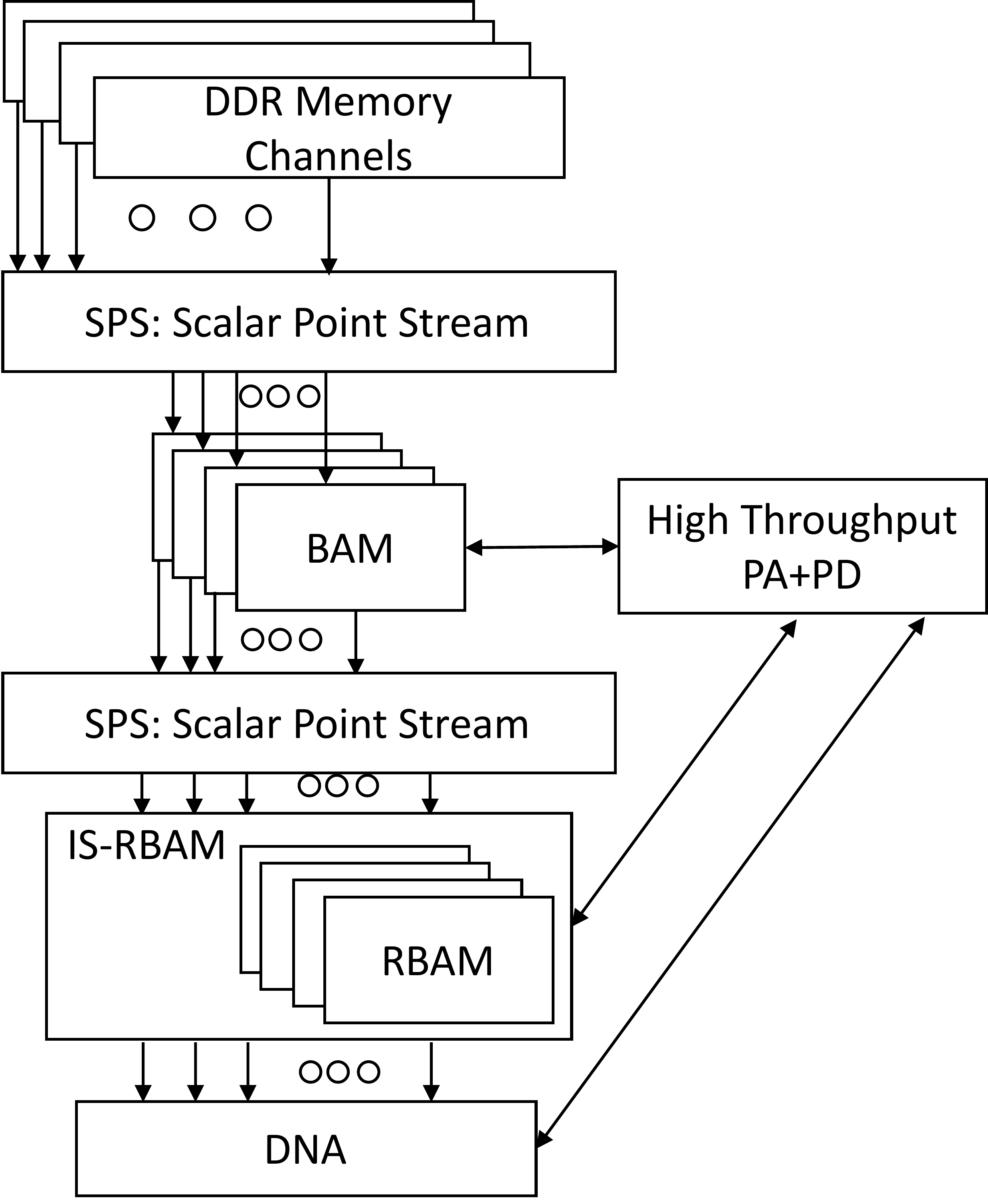} 
  \caption{SAB: Scalable Architecture for Bucket Algorithm}
  \label{fig:sab} 
  \vskip -0.3cm
\end{figure}

When looking at the MSM from a design point of view, a few important things can be inferred from the use case itself. In the context of the zk-SNARK prover, MSM is used in multiple places. However, it's crucial to emphasize that the set of elliptic curve points utilized in any given MSM operation remains constant throughout the lifetime of a given proof. So from an algorithmic point of view, only the scalars act as a primary variable input that can change with each MSM calculation request. For our implementation, we always move elliptic curve points to the main FPGA DDR memory once and consume them whenever a call to MSM is made. They are stored in DDR memory since depending on MSM size storage requirements can be in the range of tens of GBs. Our implementation target was an FPGA board based on Intel Agilex which has multiple DDR memory banks. Elliptic curve points generally need a large bit-width storage so our implementation relies on splitting and storing elliptic curve points and also scalars if needed in multiple banks. The storage layout is handled on the host side. Figure~\ref{fig:sab} gives a simplified architectural description of our implementation. The main theme of this architecture is to feed high performance point add and double unit with operands such that it's kept busy to its maximum capacity. In our implementation, we have used different design schemes to implement this unit where we tried to optimize resource consumption based on different observations from profiling. These implementation variants will be discussed later showing evolution to a final architecture that performed best. To explain this architecture first it is important to outline the philosophy behind it. The main reason to implement MSM on FPGA is to achieve the maximum possible acceleration compared to reference implementation or stated otherwise to reduce the execution time to the minimum possible. Since it's an FPGA design and when it's running on any given FPGA the whole FPGA Fabric and resources are at disposal to be utilized. Keeping all this in mind our solution uses an architecture that provides multiple knobs for scaling the implementation. These knobs can help us scale architecture to maximize throughput while trading-off FPGA resource utilization. While calculating MSM there are two core algorithms that are to be run concurrently at any given time namely {\bf Pippenger or Bucket Algorithm} and {\bf Point Add and Double Algorithm}.

The Bucket Algorithm~\cite{pippenger1976evaluation} calculates most compute-intensive parts of smaller MSMs and the "Point Double and Add Algorithm" recursively combines them into final results for large MSM.
Figure~\ref{fig:sab}, from top to bottom, starts with layered memory channels that show how it can use multiple memory channels in parallel. These parallel channels feed into a unit named Scalar-Point Stream (SPS) it fetches elliptic curve points and related scalars and forms pairs and streams them to the main compute block called Bucket Array Manager (BAM). The BAM unit does the heavy lifting and carries out the core compute that may account for generating 
 90\% or more elliptic curve point add/double operations. The BAM unit is designed to be scalable at compile/build time. It can be replicated to have as many BAM units as we can fit into the given FPGA. The BAM unit replication also takes care of forking and joining I/O and also sets priority at join and fork to avoid deadlocks. The BAM unit is the compute block that conceptually carries the smaller MSM compared to the original top-level MSM and runs in other words the \cite{pippenger1976evaluation} algorithm. The termination or accumulation part of the algorithm is generally implemented by recursive use of "Point Double and Add Calculations", in our implementation that is one of the novelties: we factored this accumulation part into another MSM and for that, we designed a compute unit called Independently Scalable Recursive Bucket Array Manager (IS-RBAM). This compute unit runs a much smaller instance of Bucket Algorithm~\cite{pippenger1976evaluation}.  The use of IS-RBAM unit reduces point double and addition operation by a large ammount that are needed for combination of small MSMS. IS-RBAM can be scaled independently without impacting the interface to other blocks. The scaling knob provided by this block helps us to optimize throughput by re-claiming resources from IS-RBAM and trying to use it for other compute units specially the BAM. The final compute unit within the SAB architecture is Double and Add (DNA) which performs the accumulation phase to arrive at the final result. It performs accumulation in a recursive fashion. First, it acts as a bit-collector for IS-RBAM and accumulates results of many smaller MSMs using multiple separate double and add phases. Finally, it combines results from these smaller MSMs into the desired top-level MSM result which is just a single elliptic curve point. The result is passed to the host side through a store unit that is not shown in the diagram for the purpose of saving space. The traffic through this store unit is negligible. For example, primary input could be in few GBs that is moved from host to DDR main memory of FPGA whereas the result is a single elliptic curve point that might take less than 100 Bytes even for curves that are defined over very wide bit-with prime fields.  

\subsection{Design of a High-Performance Point Processor}

In the previous sections we have established that the core mathematical operations on elliptic curve points during the execution of the MSM are  addition (of elliptic curve points) and doubling of a single point (when two points are determined to be the same). In this section we focus on an accelerator architecture for these operations on the FPGA platform. Both point addition \emph{(PA)} and point double \emph{(PD)} are complex compute operations. A straightforward, fully pipelined implementation of both equations would require a total of 25 \cite{hyperelliptic2007_ecc_add_formulas} instances of  254/381-bit modular multipliers alongside many adders of the same width. This steered our approach to start by having a high-performance modular multiplier implementation. Additionally, other key to this strategy was balancing the use of FPGA resources while actively seeking methods to decrease the number of necessary modular multiplier instances. Some software implementations~\cite{blst} have uncovered that point double and point add operations can share some compute and by leveraging this insight we are able to achieve full pipelining of both operations using just 18 total instances of modular multipliers. This was the result of numerous iterative implementations strategies we explored. This result is what we consider to be a state of the art implementation of elliptic curve point addition and doubling on FPGA in the form of a unified point processor.

\subsubsection{Field Arithmetic}
\label{subsec:field_arith}
To implement the equations for point addition and point doubling, the optimized building blocks we need are modular multiplication, modular addition, modular subtraction, and modular shift-by-1 operations~\cite{hyperelliptic2007_ecc_add_formulas}. This optimization predominantly hinges on the development of highly-refined adder, multiplier, square, and modulo blocks which must be tailored to facilitate efficient pipelining for large bitwidths and accept new inputs on every clock cycle. Additionally, because modular addition, modular subtraction and modular shift operations will only see inputs ($a$ and $b$) in range $[0..2N)$ (where $N$ is the modular prime), these blocks do not require a costly full modular operation due to the fact that $a + b$  or $N + a - b$ will always fall in the range $[0..2N)$ allowing us to save some FPGA resources on those operations. Starting with adders and multipliers we used common FPGA optimized implementations \cite{LanghammerAdders19, LanghammerModMult2021} and were able to take advantage of the Agilex architecture to reduce the area footprint while also increasing design frequency. 
 
In the initial stages of implementing the modulo operation, our approach incorporated the use of the multipliers, adders, and subtractors to construct a pipelined Montgomery multiplier. Our software reference implementation, $libsnark$, computed \emph{PA} and \emph{PD} in Montgomery form. Additionally, Montgomery and Barret Reduction are commonly seen as efficient hardware implementations, therefore this appeared to be a good starting point. Future developments led us to adopt a more efficient method, inspired by Ozturk's Look-Up Table (LUT)-based strategy for modular operations on FPGAs \cite{2020lowlatencymm}. Moving to LUT-based methodology significantly reduced our resource usage by allowing us to consolidate the number of multiplier blocks required for each modular multiplication within our design from three to just one. Furthermore, this LUT-based approach can be executed using either M20K blocks or DSP blocks on Agilex \cite{gribokCSAIL23}. This versatility gives an additional degree of control, enabling us to fine-tune the resource consumption of our PA and PD implementation.

\subsubsection{PA+PD Architecture}
\begin{table}
\centering
\caption{Resource Utilization for Point Adder Point Double units}
\label{table:compare_pa_pd_resources}
\begin{tabular}{c||c||c|c|c}
\hline
\bf \multirow{2}{*}{Operation} & \multirow{2}{*}{\bf Throughput}  & \multicolumn{3}{c}{\bf Resource utilization} \\
\cline{3-5}
              &                           & \bf ALMs & \bf DSP & \bf M20K \\
\hline\hline
\noalign{\vskip 1pt} 
\emph{Point Add (PA)}    & 1              & 272K & 4800 & 332 \\
\emph{Point Double (PD)} & \text{approx.} $\frac{1}{650}$  & 100.1K & 255 & 410 \\[3pt]
\hline
\end{tabular}
\vskip -0.4cm
\end{table}

Our initial implementation was based on separate \emph{PA} and \emph{PD} blocks. Software profiling revealed that the relative compute frequency of the \emph{PD} operation is less than 0.01\% that of \emph{PA}. For this reason we felt it would be inefficient to fully pipeline a standalone \emph{PD} module. The \emph{PA} module on the other hand was implemented as a fully pipelined block in RTL (allowing one \emph{PA} per cycle) with Avalon streaming interfaces to integrate into the oneAPI framework. We used a pipelined Montgomery multiplier to build our \emph{PA} module and the resulting latency was approximately 425 clock cycles, while closing timing over 700MHz.  
 
In an effort to conserve resources, our initial strategy for the \emph{PD} block involved a folded architecture constructed using the oneAPI framework that was using just a single Montgomery multiplier, a modular adder, and a modular subtractor coupled with a control block. This control block iterates the points through these functions in the required sequence of the  \emph{PD} equation. Additionally, points would only be forwarded to this block after undergoing a  \emph{PD} check within the \emph{PA} module. By adopting this approach, we traded-off some degree of parallelism for reducing FPGA resources. Table~\ref{table:compare_pa_pd_resources} shows the resource utilization for the \emph{PA} and \emph{PD} blocks. The \emph{PD} has a theoretical throughout of 1/650 clock cycles (some overhead is added by the oneAPI framework), and reports significantly lower resource utilization compared to \emph{PA}. 

\subsubsection{UDA Architecture}
\begin{figure}
    \centering
    {\scalebox{0.38}{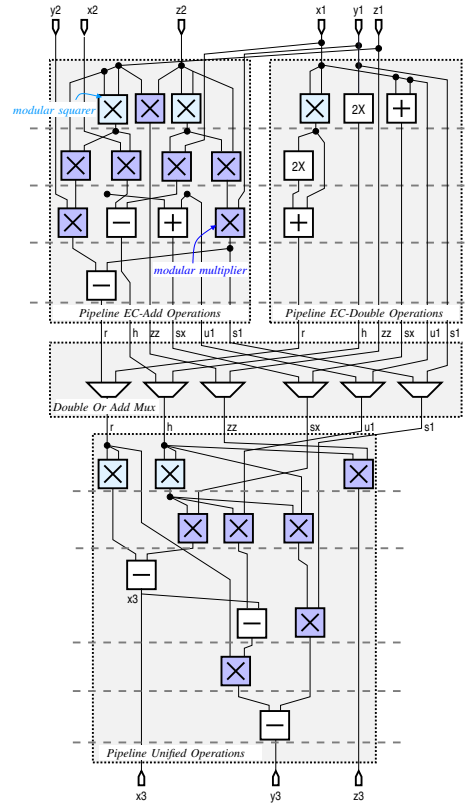}}
    \caption{Unified Double and Add Pipeline (UDA)}
    \label{fig:uda}
    \vskip -0.4cm
\end{figure}
Careful hardware experimentation with separate \emph{PA} and folded \emph{PD} units revealed that, depending on test vectors, the folded low-throughput \emph{PD} was occasionally becoming a bottleneck in the overall system performance. 
In addition, because of the area overhead that comes with high level design IP implementations, we did not save nearly as many FPGA resources as we had hoped. 
 
To remove the performance bottleneck resulting from slow point double, an implementation was devised based on a fused point add and point double calculation formula \cite{hyperelliptic2007_ecc_add_formulas} \cite{blst}. The Point Add and Point Double operations were unified into a single compute pipeline, which is called \emph{Unified Double Add (UDA)} and can be seen in \cref{fig:uda}. Initially two separate calculations are started for \emph{PD} and \emph{PA}. Four stages through the architecture, a join-mux selects intermediate results from \emph{PD} or \emph{PA} pipelines based on a \emph{PD} check. The proper intermediary variables are then forwarded onto the remaining fused portion of the architecture (with a latency of 5 stages) to calculate the final result. The pipeline is again built in RTL for a throughput of a single operation per clock cycle which can handle both \emph{PA} and  \emph{PD} calculations. 
 
This architecture was again able to close timing above 700MHz. By simply swapping out the \emph{PA+PD} core with the \emph{UDA} core, a 30\% improvement in performance was observed on the MSM. The ALM utilization was also improved by roughly 22\% as it can be observed in Table~\ref{tab:pa_variant_resource_table}.

\subsubsection{Non-Montgomery Form}
As we looked to scale-up from the 254-bit wide BN128 curve to the 381-bit wide BLS12-381 curve, it was not possible to fit the design in the target FPGA due to growth in DSP resources in relation to growth of the bitwidth. We investigated N-part Karatsuba splitting to save DSP resources but observed an undesired ALM increase as a consequence. For this reason the BLS-381 curve implementation required moving away from Montgomery domain to bypass the requirement of a \emph{Montgomery Multiplier} requiring 3 integer multipliers for each modular multiplication instance. 

We moved to implement our modular operation as a separate block (explained in Section~\ref{subsec:field_arith}) which was either M20K-based modular operation \cite{2020lowlatencymm} or DSP-based modular operation \cite{langhammer2022low_modu_mult_sergey}. Implementing both allowed for balancing resources on the FPGA and allowed us to reduce our modular multiplication block from 3 integer multipliers to 1, resulting in a significant reduction in DSP blocks along with additional reduction in ALM resources used in the \emph{UDA} architecture. 

 This final overhaul of the point processor resulted in major improvements in all aspects of the design. Our latency was reduced from 425 to 270 clock cycles while closing timing above 700MHz for BN128 (254-bitwidth) and above 600MHz for BLS12-381 (381-bitwidth). In terms of resources this resulted in a 63\% reduction of DSP resources when comparing BN128 UDA implementation in Montgomery format compared to standard number format. For ALMs we gained a 44\% reduction when comparing the original \emph{PA+PD} architecture now compared to our UDA architecture in standard format. The main cost here was a significant increase in M20K resources but this was initially a very low utilized resource which makes this an optimal tradeoff.

\begin{table}
\centering
\caption{Elliptic Curve Adder Resource Utilization}
\label{tab:pa_variant_resource_table}
\begin{tabular}{|c|c|c|c|}
\hline
\bf Design Variant & \bf ALMs Used & \bf DSP Blocks & \bf M20K \\
\hline \hline
PA+PD-254-Montgomery  & 372,700 & 5005 & 742 \\
UDA-254-Montgomery & 290,400 & 5400 & 647 \\
UDA-254-Standard & 207,000 & 1975 & 3367 \\
UDA-381-Standard & 419,000 & 4425 & 6770 \\
\hline
\end{tabular}
\vskip -0.5cm
\end{table}

\section{Performance Evaluation}
The performance evaluation for our solution is done both in absolute terms and also in a comparative fashion. The host side offloads the compute to the FPGA by sending the scalar values. Execution time measurements are recorded for different MSM lengths. In a similar fashion, measurements are recorded for the CPU and GPU implementations in order to evaluate the speedup of our proposed solution.
For the MSM implementation, the performance evaluation is done using the PCI-based FPGA accelerator card \cite{Bittware_Agilex}, which is  built around the Intel Agilex FPGA AGFB027R25A2E2V device.

\subsection{Methodology}
In order to measure the performance of an FPGA-based solution, test vectors for different sizes of MSM are generated by using \emph{libsnark} \cite{libsnark} and are stored in text files. The host-side application (written in C++ and running on an Intel Xeon platform) loads these test vectors in the FPGA main memory, triggering the start of the hardware execution. Upon completion of the MSM calculations, the results are  stored back in the same FPGA main memory. The host recovers the results and compares them against the golden references.
During the computation, the measurements are taken for FPGA execution-time using a high-resolution stopwatch on the host side.
A very similar strategy is employed for taking measurements in the case of the CPU implementation: the software code is instrumented with stopwatch triggers and logging capabilities. The GPU implementation is again software-based so measurements are taken in a similar fashion to the CPU implementation. In our approach to GPU profiling, we are reliant on open-source libraries and hardware supported by Cloud Service Providers due to the absence of an in-house implementation. This led us to the Bellperson library \cite{bellperson} as a suitable benchmark for our comparison of the BLS12-381 MSM. This library was accessible via an Amazon EC2 instance and is compatible with GPUs based on the Turing architecture or newer. For our analysis, we utilized the g4dn.16xlarge EC2 instance, which is equipped with an NVIDIA T4 GPU. 

\subsection{CPU Performance: Software Profiling Results}
For performance analysis on the CPU, \emph{libsnark} \cite{libsnark} is used both as a golden reference for functional verification, as well as the performance baseline. This is an open-source C++ library for zk-SNARK applications that provides several APIs that can be used to construct circuits and set-up \emph{zk-SNARKs} -- that can be later used by the \emph{verifier} and the \emph{prover}.
Since we are interested in benchmarking performance of both the BN128 and BLS12-381 elliptic-curve implementations, we make use of the open-source \emph{Clearmatics} implementation \cite{clearmatics2023libsnark} that is forked from \emph{libsnark} and adds the BLS12-381 support.

\begin{table}
\centering
\caption{Platform Details used for Profiling}
\label{table:platform_details_fpga_gpu}
\begin{tabular}{l||c|c}
\hline
 & \textbf{CPU \& FPGA} & \textbf{GPU} \\ 
 \hline\hline
CPU         & Xeon Silver 4310   & Xeon Platinum 8259CL\\
CPU Frequency & 2.10GHz & 2.50GHz \\
CPU Cores & 48 & 64 \\ 
Total RAM & 188GB & 248GB \\
OS Version & CentOS Stream 8 & Ubuntu 20.04 LTS \\ \hline
\end{tabular}
\vskip -0.5cm
\end{table}

\begin{figure}
    \centering
    \includegraphics[width=0.40\textwidth]{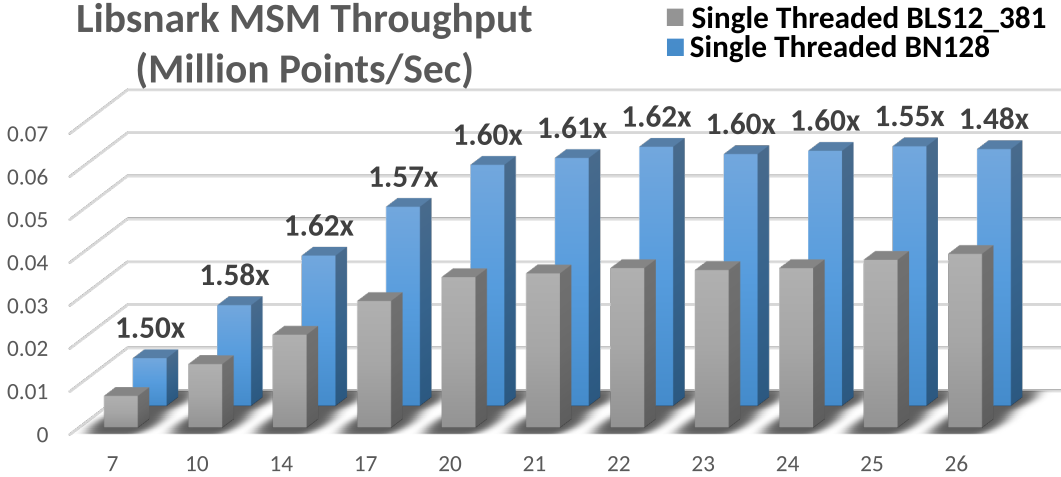}
    \caption{MSM Throughput running on CPU}
    \label{fig:libsnark_throughput}
\end{figure}

The details of the platforms used for performance profiling are shown in Table~\ref{table:platform_details_fpga_gpu}. 
To profile MSM and generate golden test vectors, relevant APIs are identified and instrumented using a simple stopwatch implemented in C++ and logging functions. The profiling results are presented in Figure \ref{fig:libsnark_throughput}. The performance is reported in terms of (millions of) MSM points processed per second \emph{(M-MSM-PPS)}, which is calculated by normalizing the execution time by MSM size (number of points). The results shown in Figure~\ref{fig:libsnark_throughput} show that, in the case of a single-threaded CPU implementation, throughput flattens as the MSM size increases, with the highest throughput being achieved for very small size MSM. In absolute terms, the throughput for BN128 \cite{bn128_curve_standard_draft} peaks at roughly 0.06 \emph{M-MSM-PPS} and for BLS12-381 curve  \cite{BarretoNaehrig2006_bls12_381_curve_ref_bn128_ref} its the throughput peaks at 0.04 \emph{M-MSM-PPS}. The throughput for BLS12-381 is lower than for BN128 simply due to the wider bitwidths associated with operations on this elliptic curve.

\subsection{FPGA Implementation Performance Evaluation}
Our FPGA implementation for MSM has gone through significant evolution while targeting performance improvements and resource utilization reduction. The overall implementation is a hardware-software split where the CPU sees an FPGA-based MSM accelerator as an offload compute engine. The host interface and information exchange are very simple it uses a load-store mechanism for I/O. Each load or store triggers a data movement from the host main memory to the FPGA main memory. The data transfer completion signals the start of computation on the FPGA side or the availability of the final result on the host side. Rigorous experimentation and benchmarking are carried out for the FPGA implementation and results are presented in a format that allows us to compare these with the CPU and GPU implementations.
\subsubsection{System Level FPGA Resource Utilization}
\begin{table}
\centering
\caption{System Level Resource Utilization}
\label{tab:system_level_resource_util}
\begin{tabular}{|l|c|c|c|}
\hline
\bf Design Variant & \bf ALMs & \bf DSP Blocks & \bf M20K \\
\hline \hline
BN128 PAPD-Montgomery(S=2)  & 715,603 & 5005 & 4642 \\
BN128 UDA-Standard(S=2) & 571,408  & 1975 & 6501 \\
BN128 UDA-Standard(S=1) & 537,348 & 1975 & 5616 \\
\hline\hline
BLS12-381 UDA-Standard(S=2) & 831,972 & 4425 & 10,973 \\
BLS12-381 UDA-Standard(S=1) & 770,561 & 4425 & 9662 \\
\hline
\end{tabular}
\end{table}

The resource utilization for various builds are reported in Table~\ref{tab:system_level_resource_util}. The FPGA device that we are using has total 912,800 ALMs and for BLS12-381 curve with scaling=2 the ALM utilization peaks at 91\% which is very close to FPGA capacity ceiling. For this configuration achieved fmax was 351MHz. For other build variations fmax was in the range of 334-367MHz. Table~\ref{tab:system_level_resource_util} also provides insights into architectural evolution and resulting resource utilization effects. Switching to UDA (S=2) architecture from PAPD provides 21\% reduction in ALMs and a 60\% reduction in DSPs whereas M20K/BRAMs utilization goes up by 40\%. In a nutshell, switching from PAPD to UDA provides significant reduction in ALMs which is generally the most contested resource and also the potential bottleneck creating challenges while in the process of placement and routing. The interesting thing to note is the effect of scaling on resource utilization, for BLS12-381 the upward delta is 7\% on ALMs and 13\% on BRAMs whereas the DSPs stay the same since both configurations use single UDA instance underneath. Importantly latter section will show that with scaling throughput goes up linearly.   

\subsubsection{Throughput Measurements}
\begin{figure} 
  \centering
  \includegraphics[width=.85\linewidth]{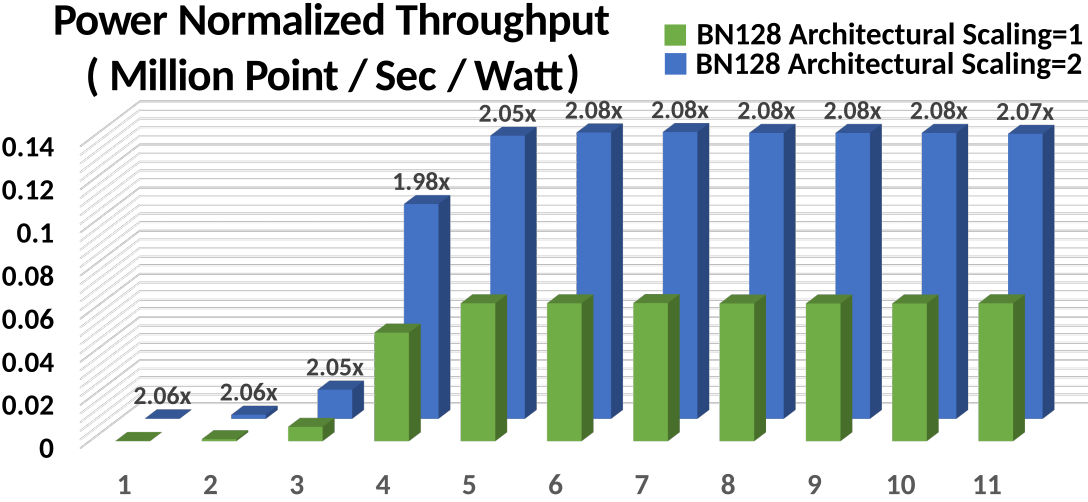} 
  \caption{FPGA Power Normalized Performance for BN128 Curve}
  \label{fig:fpga_normalized_throughput_bn128_s1_s2} 
\end{figure}

\begin{figure}
  \centering
\includegraphics[width=.95\linewidth]{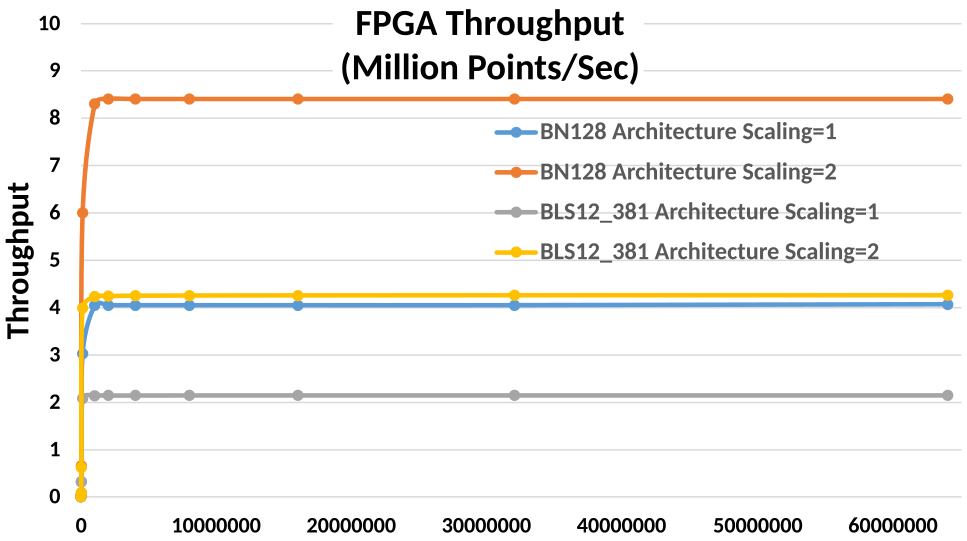} 
    \caption{FPGA Performance Across Curve and Scaling }
  \label{fig:fpga_throughput_for_bn_bls_s1_s2} 
  \vskip -0.4cm
\end{figure}

The MSM FPGA implementation performance results for the two different curves are presented in Figure~\ref{fig:fpga_throughput_for_bn_bls_s1_s2}. The results are also given for two architectural variations for each curve. The first thing to note is that throughput peaks with respect to increasing MSM size relatively early: MSM sizes with tens of thousands of points will execute at maximum throughput. The lower throughput observed for small-size MSM is caused by the host-device communication and control overhead.

The performance of BN128~\cite{bn128_curve_standard_draft} and BLS12-381~\cite{BarretoNaehrig2006_bls12_381_curve_ref_bn128_ref} curves is simply different due to compute complexity, the performance of BN128 is almost 2x compared to BLS12-381. The scaling of architecture almost linearly improves throughput for each curve. The scaling is currently limited only by the availability of resources on FPGA and its impact on maximum clock frequency.

\begin{figure}
  \centering
  \includegraphics[width=.85\linewidth]{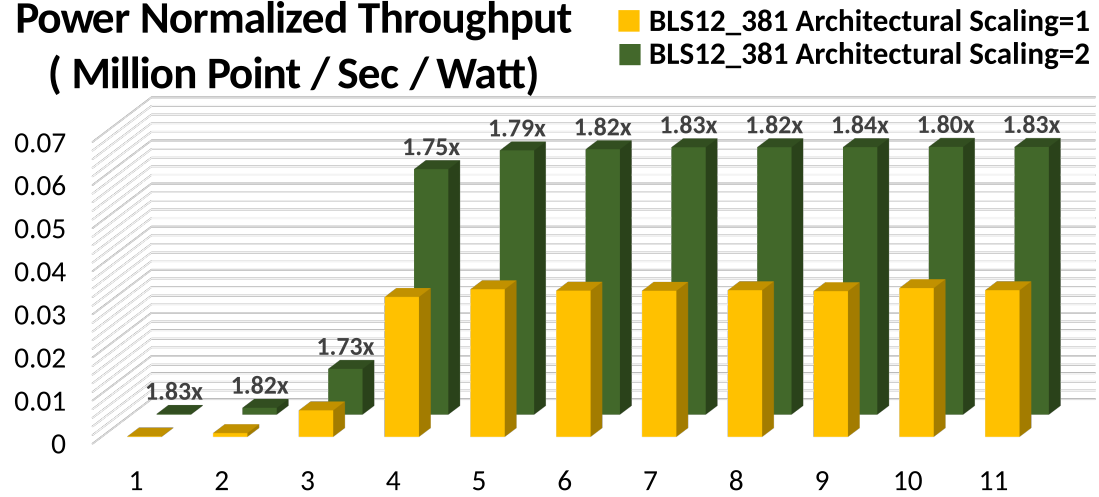} 
  \caption{FPGA Power Normalized Performance for BLS12-381 Curve}
  \label{fig:fpga_normalized_throughput_bls12_381_s1_s2} 
\end{figure}

\begin{figure}
  \centering
  \includegraphics[width=.83\linewidth]{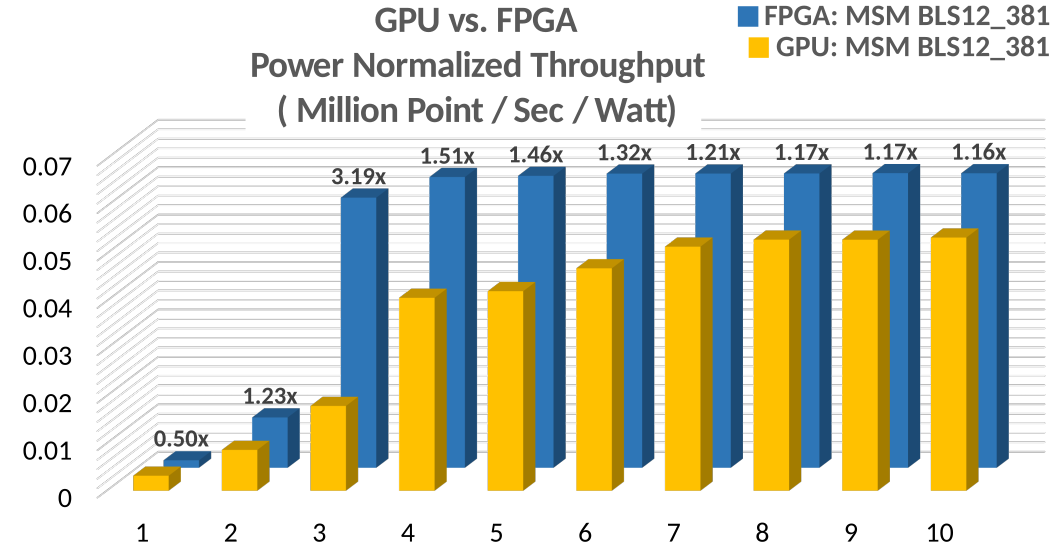} 
  \caption{FPGA vs. GPU Comparison for Normalized Throughput}
  \label{fig:normalized_fpga_gpu_througput_comparison} 
  \vskip -0.2cm
\end{figure}

Figure~\ref{fig:fpga_normalized_throughput_bn128_s1_s2} gives power normalized throughout for two different architectural scaling factors. Its for BN128 curve and in terms of throughput it highlights similar performance pattern where performance peaks early for MSM size but the most interesting observation is about the effect of architectural scaling on power efficiency. The measurements clearly show that higher scaling factor of 2 is almost giving a power efficiency that is 2x better compared to scaling factor 1. Its very promising result in terms of MSM FPGA implementation power consumption. Where higher scaling can achieve higher throughput with little increase in power consumption. Hence linear throughput scaling and also almost linear power normalized throughput scaling as well. We have evaluated this trend for both the curves namely BN128 and BLS12-381 as shown in Figure~\ref{fig:fpga_normalized_throughput_bn128_s1_s2} and Figure~\ref{fig:fpga_normalized_throughput_bls12_381_s1_s2}. Our evaluation is possible for only two scaling factors because of the resources available on the FPGA but it is evident that this trend will hold for other scaling factors, and might be part of our future work when other larger devices are available for experimentation.

\subsubsection{Power Measurements}
\begin{table}
\centering
\caption{FPGA Power Measurements for MSM Design}
\label{tab:power_standby_active_msm}
\begin{tabular}{|l|c|c|}
\hline
\bf Design Variant & \bf Standby Power & \bf Active Power \\
\hline \hline
oneAPI BSP Only      & 17.25 & N/A  \\
BN128 PAPD (Scale=1) & 44.6  & 72.7 \\
BN128 UDA  (Scale=1) & 42.6  & 58.0 \\
BN128 UDA  (Scale=2) & 44.7  & 63.5 \\
BLS12-381 UDA (Scale=1) & 48.8  & 63.1 \\
BLS12-381 UDA (Scale=2) & 50.4  & 68.6 \\
\hline
\end{tabular}
\vskip -0.3cm
\end{table}
Table~\ref{tab:power_standby_active_msm} gives two different type power measurements for various implementations while computing MSM of size 64M points. Standby power measurement readings are taken when FPGA bit-streams are loaded but no input is being processed meaning the compute kernels are idling. The only different type of reading for power is the BSP only power measurement where FPGA has a minimal bit-stream configuration that carries the shell that allows FPGA platform to be ready for communication with HOST device. The BSP has many functions but one of the important ones is to load a full bit-stream that carries the image that configures user design carrying all the compute kernels. In our case all the hardware kernels that we have to calculate MSM and perform data movements. The board power consumption is smallest with BSP since it uses minimum logic. Once the actual design is configured the standby power consumption goes up and is proportionally related to logic utilization that can be verified by referring to Table~\ref{tab:system_level_resource_util}. For example the UDA based architecture uses less logic compared to PAPD hence less standby power. The other interesting and very promising effect to note is that power consumption doesn't go up linearly with scaling, meaning scaling help greatly in improving performance/watt. By comparing numbers in Tables~\ref{tab:system_level_resource_util} and \ref{tab:power_standby_active_msm} it can be observed that power consumption correlates well with total resource utilization.

\subsubsection{Performance Comparison: CPU vs. FPGA vs. GPU}
Table~\ref{tab:compare_exec_time_cpu_gpu_fpga} gives execution time measurements clearly showing that FPGA out performs CPU( utilizing multiple core libsnark implementation while using OpenMP) and achieves acceleration factors that range from 7-124x for various MSM sizes. Specially for large size MSMs FPGA performance gains is always larger than 100x. FPGA performed at par or better compared to GPU used for benchmarking with maximum performance gain of 3x. For larger MSM sizes FPGA performed between 15-30\% better comapred to GPU. For GPU comparison power normalized throughput plot is also give in Figure~\ref{fig:normalized_fpga_gpu_througput_comparison} and it shows that FPGA has reasonable power advantage of GPU always outperforming by 16-51\% for large size MSMs,even though a low power GPU was used for benchmarking. Lastly, Table~\ref{tab:comapre_cpu_fpga_gpu_perf} gives absolute power measurements and execution time for 64M point MSM. It shows that FPGA uses 11\% less power and has 10\% better throughput compared to the GPU. For CPU similar power numbers can't be produced easily hence not available.

\begin{table}
\centering
\caption{Execution Time Comparison for BL12\_381 Curve}
\label{tab:compare_exec_time_cpu_gpu_fpga}
\begin{tabular}{|r|c|c|c|c|c|}
\hline
\multicolumn{1}{|c|}{} & \multicolumn{3}{c|}{Execution Time (s)} & \multicolumn{2}{c|}{FPGA Perf. Gain (s)} \\
\hline
\textbf{MSM Size} & \textbf{CPU} & \textbf{GPU} & \textbf{FPGA} & \textbf{xCPU} & \textbf{xGPU} \\
\hline
1,000 & 0.07 & 0.01        & 0.01    & 7x    & 1.00x\\
10,000 & 0.46 & 0.02       & 0.02    & 23x   & 1.00x\\
100,000 & 3.39 & 0.09      & 0.03    & 113x  & 3.00x\\
1,000,000 & 29.92 & 0.36    & 0.24    & 124x  & 1.50x\\
2,000,000 & 58.39 & 0.68    & 0.47    & 124x  & 1.45x\\
4,000,000 & 112.90 & 1.21   & 0.94    & 120x  & 1.29x\\
8,000,000 & 228.61 & 2.21   & 1.88    & 121x  & 1.18x\\
16,000,000 & 451.70 & 4.28  & 3.76    & 120x  & 1.14x\\
32,000,000 & 858.78 & 8.63  & 7.51    & 114x  & 1.15\\
64,000,000 & 1658.88 & 17.10& 15.03   & 110x  & 1.14x\\
\hline
\end{tabular}
\end{table}

\begin{table}[t]
\centering
\caption{Performance comparison between CPU, GPU and FPGA }
\label{tab:comapre_cpu_fpga_gpu_perf}
\begin{tabular}{|l||c|c|c|c|}
\hline
\multirow{2}{*}{\bf Device} & \multicolumn{2}{c|}{\bf Execution Time (s)} & \multicolumn{2}{c|}{\bf Power Consumption (W)} \\
\cline{2-5}
& BN128 & BL12\_381 & BN128 & BL12\_381 \\
\hline \hline
CPU & 1123 & 1658 & NA & NA \\
\hline
GPU & NA & 17.1 & NA & 70 \\
\hline
FPGA & 7.6 & 15 & 68 & 63 \\
\hline
\end{tabular}
\vskip -0.4cm
\end{table}

\section{Conclusion and Future Work}
The field of Zero-Knowledge Proofs (ZKPs) is currently experiencing a lot of innovation, which is leading to their application in numerous critical domains. Given the computationally intensive nature of ZKPs, for all practical applications they should be accelerated using different targets. Our proposed FPGA architecture combined with our optimized low-level  IPs allowed us to achieve significant acceleration for multi-scalar multiplication compared to reference software library. The benchmarking against GPU for performance and performance/watt proves the advantage our FPGA implementation can provide. 

In future we plan to accelerate NTT/INTT on FPGAs and also adapt our implementation to G2 type MSM. In terms of architecture and implementation strategies there is still lot of room to extract more performance by enabling more memory banks, trade-off resources between different architectural blocks based on compute requirements and also to optimize low level modular arithmetic IP giving us more room for scaling. Given the nature of problem there is significant motivation to experiment with Multi-FPGA implementation.

\IEEEpeerreviewmaketitle
\balance
\bibliographystyle{IEEEtran}
\bibliography{if-ZKP}
\end{document}